\documentclass{Interspeech}

\interspeechcameraready 

\title{Neural Spectral Band Generation for Audio Coding}

\author[affiliation={1}]{Woongjib}{Choi}
\author[affiliation={1}]{Byeong Hyeon}{Kim}
\author[affiliation={1}]{Hyungseob}{Lim}
\author[affiliation={2}]{Inseon}{Jang}
\author[affiliation={1}]{Hong-Goo}{Kang}

\affiliation{Department of Electrical and Electronic Engineering}{Yonsei University}{Seoul, South Korea}
\affiliation{Electronics and Telecommunications Research Institute}{Daejeon}{South Korea}
\email{woongzip1@dsp.yonsei.ac.kr, bhkim98@dsp.yonsei.ac.kr, hyungseob.lim@dsp.yonsei.ac.kr, jinsn@etri.re.kr, hgkang@yonsei.ac.kr}
\email{\{woongzip1, bhkim98, hyungseob.lim\}@dsp.yonsei.ac.kr,
\\ jinsn@etri.re.kr, hgkang@yonsei.ac.kr
}
    

\usepackage{caption}
\usepackage{subcaption}
\usepackage{comment}
\usepackage{cite}
\usepackage{multirow}
\begin{document}

\maketitle

\begin{abstract}
Spectral band replication (SBR) enables bit-efficient coding by generating high-frequency bands from the low-frequency ones.
However, it only utilizes coarse spectral features upon a subband-wise signal replication, limiting adaptability to diverse acoustic signals. 
In this paper, we explore the efficacy of a deep neural network (DNN)-based generative approach for coding the high-frequency bands, which we call neural spectral band generation (n-SBG). 
Specifically, we propose a DNN-based encoder-decoder structure to extract and quantize the side information related to the high-frequency components and generate the components given both the side information and the decoded core-band signals. 
The whole coding pipeline is optimized with generative adversarial criteria to enable the generation of perceptually plausible sound. 
From experiments using AAC as the core codec, we show that the proposed method achieves a better perceptual quality than HE-AAC-v1 with much less side information.
% ===> 1003 characters
% HE AAC v1 -> HE AAC ?
\end{abstract}

\renewcommand{\thefootnote}{}
\footnotetext{
    This work was supported by Electronics and Telecommunications Research Institute (ETRI) grant funded by the Korean government. [25ZC1100, The research of the basic media$\cdot$contents technologies]}
\renewcommand{\thefootnote}{\arabic{footnote}}
\section{Introduction}

% Kang: bitrate vs. bit-rate 통일 할 것.
% ===> bit-rate

% Goal of audio coding (bit-rate)
The primary objective of audio coding (or audio compression) is to reduce the amount of data needed to represent audio signals while preserving the quality of the decoded output~\cite{book_bosi, painter2000perceptual}. 
% Audio coding types
Audio coding schemes can be classified as either lossless or lossy.
% Goal of lossless coding
Lossless coding preserves the original signal exactly, whereas lossy coding achieves higher compression by permitting some quality loss.
%--------------
% while lossy coding prioritizes reproducing a signal perceptually indistinguishable as much as possible from the input signal.
% Lossless coding ensures the exact reconstruction of the original audio signal by guaranteeing identical bit-level representations.
% However, this precision often imposes significant constraints on the achievable compression ratio, as the amount of information required to represent the signal is inherently limited by these stringent requirements.
%--------------
% Goal of lossy coding
%--------------
% In contrast, lossy coding prioritizes perceived transparency over exact reconstruction, aiming to reproduce a signal that is perceptually indistinguishable from the original audio signal.
%--------------
% advantage of lossy coding
%--------------
% By relaxing the constraint of bitwise identical reconstruction, 
% By removing or simplifying audio components that are less critical to human perception, 
%--------------
% On the other hand, lossy coding can achieve a significantly higher compression ratio than lossless coding while maintaining satisfactory audio quality.
%--------------
% Lossy coding method
% Perceptual audio coding
In particular, perceptual audio coding~\cite{painter2000perceptual,johnston1988estimation}, a form of lossy compression, aims to maximize the data efficiency while maintaining perceived audio quality as much as possible.
% psychoacoustics
These methods exploit psychoacoustics~\cite{fastl2006psychoacoustics}---studies on the relationship between acoustic stimuli and human auditory perception---to determine the optimal bit allocation over frequency bins under a bit-budget restriction. 
By employing psychoacoustic models (PAMs) to assess the perceptual saliency of audio components, perceptual audio codecs assign fewer bits to less perceptually significant components, thereby achieving a higher compression ratio without noticeably degrading quality.
% ------
%Perceptual codecs operate by utilizing a psychoacoustic model (PAM), which analyzes the input audio signal to analyze the relevance of audio components to human perception.
% This model accounts for key psychoacoustic principles, including masking effects --- where louder sounds can obscure quieter ones --- and the varying sensitivity of human hearings across different frequency ranges.
% such as masking effects, where louder sounds can render quieter sounds inaudible, and the varying sensitivity of human hearings across different frequency ranges.
% Based on these principles, perceptual codecs as MP3 (MPEG Audio Layer III) \cite{mp3_ISO11172-3, mpeg2_ISO13818-3} and AAC (Advanced Audio Coding) \cite{aac_bosi1997iso} selectively allocate bits to frequency components in the transform domain that contribute the most to perceived audio quality.
% Under moderate bit-rate conditions, these codecs produce sound that is nearly indistinguishable from the original audio signal.
% By achieving high compression ratios while maintaining audio quality, they are well-suited for applications such as music streaming and telecommunication.
% ------
% Loss of high frequency

At low bit-rates, many perceptual audio codecs~\cite{mp3_ISO11172-3, 
% mpeg2_ISO13818-3, 
aac_bosi1997iso} opt to totally discard some spectral components above a certain frequency, prioritizing the encoding of lower-frequency content. While this strategy efficiently reduces bit-rate requirements, the resulting bandwidth limitation can lead to muffled sound quality~
% \cite{book_larsen2005audio, pp145}.
\cite[Chapter~5]{book_larsen2005audio}.
To address these issues, many perceptual codecs such as HE-AAC~\cite{mpeg4_HEAAC_ISO}, MP3Pro~\cite{mp3pro_ziegler2002enhancing} and Opus \cite{opus} incorporate spectral band replication (SBR)~\cite{sbr_dietz2002spectral} (or a similar method), a technique designed to perceptually encode the high-frequency components using the existing core-band signals in an efficient manner.
% Instead of directly encoding all spectral components, SBR extracts key parameters that facilitate the synthesis of plausible high-frequency components while the perceptual (core) codec encodes only the low-frequency components.
% SBR encoder
In the encoding stage, the input audio signal is analyzed by a filterbank and the key encoding parameters such as spectral envelope information and noise level estimates are extracted by a SBR encoder to be used for reconstructing the high-frequency components from the low-frequency ones. % 수정
% which characterizes the noisier parts of the spectrum that cannot be replicated using harmonic content alone. 
These parameters are then quantized and transmitted to the decoder.
% SBR decoder
In the decoding stage, the core codec output is separated into subband signals, and a SBR decoder reconstructs high-frequency components by replicating low-frequency subband signals into the high-frequency range.
The replicated subband signals are then adjusted based on the transmitted parameters and further enhanced with sinusoids or noise if needed.
% SBR + Codecs

% \begin{figure*}[t]
%   \centering
%   \includegraphics[width=18.2cm]{figures/fig1.jpg}
%   % \vspace{-0.1cm}
%   \caption{Overview of the framework}
%   \label{fig:framework}
%   \vspace{-0.1cm}
% \end{figure*}

% Kang: 그림을 이렇게 그리면, reviewer들이 오해할 수 있단다. 확실하게 표현하기 위해서는 현재 그림에서 Decoder 부분으로 표시된 Core codec decoder와 SBG Decoder의 analysis 구조는 Encoder에도 동일하게 표현되어야한단다. 이 때 analysis 변수는 Encoder/Decoder에서 동일하게 적용되는 Siamese network 형태로 표시하도록 하거라.
% learnable parameters, fixed module 구분하기
\begin{figure}[t]
% \flushleft 
% \hfill
% \vspace{0.1cm}
\centering
\includegraphics[width=\columnwidth*1]{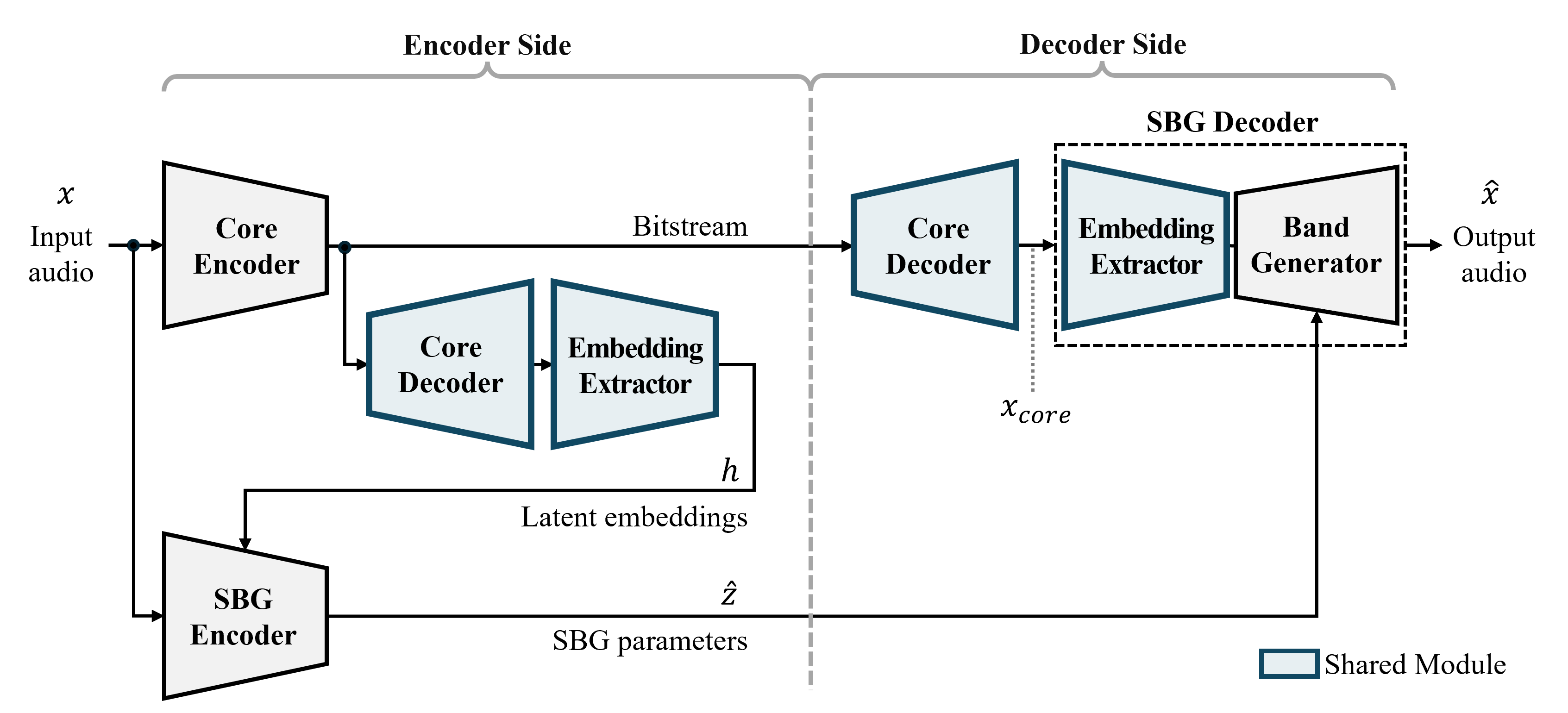}
  \caption{An overview of neural Spectral Band Generation (n-SBG). The SBG encoder extracts quantized parameters from the input audio, while the SBG decoder generates high-frequency subbands based on the transmitted parameters and coded low-frequency bands. 
  % The output embedding from the SBG decoder's bottleneck is used as a conditional input for the encoder.
  The core decoder and embedding extractor modules are shared across the encoder and decoder sides,
  while the core encoder and core decoder remain fixed and not trained.
  }
  \label{fig:framework}
  \vspace{-0.4cm}
\end{figure}

% Appliny BWE to codec outputs
% The SBR shares a similar objective with the bandwidth extension (BWE) task in a sense that both aim to reconstruct missing high-frequency components from a band-limited signal.
% Inspired by recent advancements in deep neural network (DNN)-based BWE~\cite{lee2021nuwave, li2021StreamingSEANEt,   lu2024apbwe}, one may expect to replace the SBR with a DNN-based BWE model in the audio coding pipeline.
% However, this straightforward integration may be unsuitable for general audio and music signals, as their high-frequency components exhibit complex distributions and weak correlation with lower frequencies~\cite{sbr_ekstrand2002bandwidth}.
% Unlike BWE tasks, where the missing high-frequency content must be inferred without prior information, audio coding has access to the full-band signal at the encoder. 
% This fundamental difference allows for the extraction of extra information that guides high-frequency reconstruction more accurately.

% ----------------- 05.28 --------------
The SBR is a well-established parametric approach to audio bandwidth extension (BWE), a task of reconstructing missing high-frequency components from band-limited signals. Specifically in the case of SBR, low-frequency band information along with encoded side information is utilized to generate high-frequency spectral content.
% Audio bandwidth extension (BWE) is the task of reconstructing missing high-frequency components from a band-limited signal. The SBR is a well-established parametric approach to BWE, which utilizes low-frequency band information along with encoded side information to generate high-frequency spectral content.
In parallel, numerous DNN-based BWE \cite{han2022nuwave2, liu2024audiosr, li2021StreamingSEANEt, lu2024apbwe} approaches have emerged,
with the aim of generating missing high-frequency spectra from low-band inputs (i.e., blind BWE).
% With the aim to generate missing high-frequency spectra, these techniques have often demonstrated strong effectiveness on speech signals.
Building upon these approaches, one may expect to replace the SBR with a DNN-based BWE model in the audio coding pipeline.
However, this straightforward integration may be unsuitable for general audio signals, as the correlation between the low-band and the high-band characteristics varies depending on the content~\cite{sbr_ekstrand2002bandwidth}.
Unlike these blind approaches, where the missing high-frequency content must be inferred without prior information, audio coding has access to the full-band signal at the encoder \cite[Chapter~5]{book_larsen2005audio}.
This fundamental difference allows for the extraction of extra information that guides high-frequency reconstruction more accurately.

% Proposal
Inspired by SBR, we propose \textbf{neural Spectral Band Generation (n-SBG)} as a new framework to integrate DNN-based BWE models into traditional audio coding pipelines for more efficient high-frequency restoration.
% Specific description
Figure \ref{fig:framework} provides an overview of the proposed n-SBG framework.
During encoding, a feature map is extracted and quantized from the original full-band audio, then transmitted as SBG parameters.
The decoder utilizes these parameters together with the core codec's band-limited output to generate the desired high-frequency bands.
% -----------------------
% The SBG encoder first extracts a feature map from the short-time Fourier transform (STFT) coefficients of the original audio signal. This feature map is then quantized and transmitted to the decoder side.
%The extracted side information is then discretized through a residual vector quantization (RVQ) layer, allowing it to be compressed for transmission or storage.
% After the core codec reconstructs a band-limited signal, the SBG decoder synthesizes the missing high-frequency components using the received quantized parameters.
% Finally, the generated high-frequency bands are integrated into the core codec's output, producing the full-band audio signal.
% SBG decoder reconstructs high-frequency components by modulating its internal activations with the side information. 
% -----------------------
%The entire system adopts a fully convolutional architecture in which the encoder, quantizer, and decoder are jointly trained in an end-to-end manner. 
%This design ensures that each module is co-optimized, enabling efficient high-frequency restoration even under low-bit-rate conditions.
% Contributions
% ------ Can be modified
The key contributions of this research can be summarized as follows:
\begin{itemize}
\item We propose a novel framework for parametric coding of high-frequency bands at low bit-rate conditions, replacing the rule-based SBR with a DNN-based encoder-decoder architecture.
% \item 
% We systematically compare feature extraction strategies for our SBG encoder, exploring STFT-domain parameters, core codec information to identify which aspects are most beneficial for high-frequency synthesis.
\item 
Our work lays the foundation for a new paradigm in high-frequency coding by leveraging information from core codec outputs for more efficient feature extraction and accurate reconstruction.
% Our framework can be seamlessly integrated into existing audio coding pipelines, serving as an end-to-end pre- and post-processing module to enhance overall coding efficiency.
\end{itemize}

\section{Related Works}
\subsection{Bandwidth Extension}
\vspace{-2.5pt}
% DNN based BWE
% Recent advances in deep learning have led to various DNN-based BWE approaches~\cite{lee2021nuwave, li2021StreamingSEANEt, lu2024apbwe}, which generate missing high-frequency components based on distributions learned from full-band audio data.
% % Blind BWE
% These approaches are commonly referred to as blind BWE~\cite[Chapter~5]{book_larsen2005audio}, as they do not rely on any prior knowledge about the lost high-frequency components in a specific target signal.
% % effectiveness for speech
% The blind BWE has been found to be particularly effective for speech signals~\cite{lu2024apbwe}, where high-frequency patterns are more predictable compared to those in music and general audio.

DNN-based blind BWE approaches typically use generative models such as diffusion models\cite{han2022nuwave2, liu2024audiosr} or generative adversarial networks (GAN) \cite{li2021StreamingSEANEt, lu2024apbwe} to estimate high-frequency content from low-frequency inputs, based on distributions learned from full-band audio data.
These methods have been found to be particularly effective for speech signals~\cite{lu2024apbwe}, as high-frequency patterns in speech are more predictable than those in music and general audio, 
but their performance for general audio has not yet been fully validated. 

\subsection{Neural Audio Codecs}
\vspace{-2.5pt}
Neural audio codecs (NACs) compress signals through an encoder-quantizer-decoder pipeline, often incorporating residual vector quantization (RVQ) to discretize latent embeddings from the encoder.
% SoundStream, Encodec, DAC
Representative examples such as SoundStream~\cite{zeghidour2021soundstream}, EnCodec~\cite{defossez2022encodec}, and DAC~\cite{kumar2024DAC} employ convolutional autoencoders to process time-domain audio signals.
% Adversarial training
To further improve perceptual quality, these methods often incorporate adversarial training~ with frequency-domain discriminators.
% Related work (codec + BWE)
Recently, a blind BWE module was integrated into a NAC to reduce bit-rate~\cite{ai2024low_ap}, primarily focusing on speech signals.
To generalize across diverse audio signals, our approach transmits additional parameters for high-frequency generation through a dedicated encoder module.

\subsection{Post-Processing with Auxiliary Information}
\vspace{-2.5pt}
While many works aim to enhance degraded codec outputs solely based on the decoded signal itself~\cite{biswas2020audioenhance_c,deng2020lstmrnnaudio, gupta2022dnn_mdct}, some post-processing modules incorporate features obtained from the input signal before encoding to improve the performance of the enhancement.
% For example, \cite{hwang2021nnbasedenhance_c} and \cite{lin2022codec2enhance_c} extract information from the uncompressed signal and use it as additional conditioning to refine the reconstructed signal.
% However, these methods primarily focus on speech signals with low sampling rates.
% 어떤식으로 추가정보 추출?
% CNN / Frequency domain
% For instance, \cite{hwang2021nnbasedenhance_c} and \cite{lin2022codec2enhance_c} propose extracting and transmitting DNN-based side information to refine the decoded output of the core codec. % DNN-based side information => 풀어쓰기 
For instance, \cite{hwang2021nnbasedenhance_c} and \cite{lin2022codec2enhance_c} extract quantized features from the frequency-domain representation of the input signal using convolutional networks, which are then provided for the enhancement.
However, these methods primarily focus on speech signals with low sampling rates.

% \begin{figure*}[t]
%     \centering
%     \begin{subfigure}[b]{0.47\textwidth}
%         \centering
%         \includegraphics
%         % [width=\linewidth]
%         [scale=0.47]
%         % {figures/Fig2-a_core_cond.png}
%         {figures/fig2a-v2.jpg}
%         \caption{}
%         \label{fig:fig2a}
%     \end{subfigure}
%     \hfill
%     \begin{subfigure}[b]{0.47\textwidth}
%         \centering
%         \includegraphics
%         [scale=0.47]
%         % [width=\linewidth]
%         % {figures/Fig2-b_core_cond.png}
%         {figures/fig2b-v2.jpg}
%         \vspace{9pt}
%         \caption{}
%         \label{fig:fig2b}
%     \end{subfigure}
%     \vspace{-10pt}
%     \caption{Detailed descriptions of the encoder and the decoder architecture: (a) SBG encoder architecture and details of the feature encoder and the projection layer, (b) SBG decoder architecture}
%     \label{fig2}
%     \vspace{-0.3cm}
% \end{figure*}
% % learnable, fixed 구분 (가능하면 색으로), label 달아주기
% % SBG Encoder: Residual Unit -> Residual Stage
% % FiLM -> TFiLM

\begin{figure*}[t]
    \centering
    \includegraphics
     [trim={0 6pt 0 10pt}, clip, width=\linewidth]
    % [scale=0.47]
    {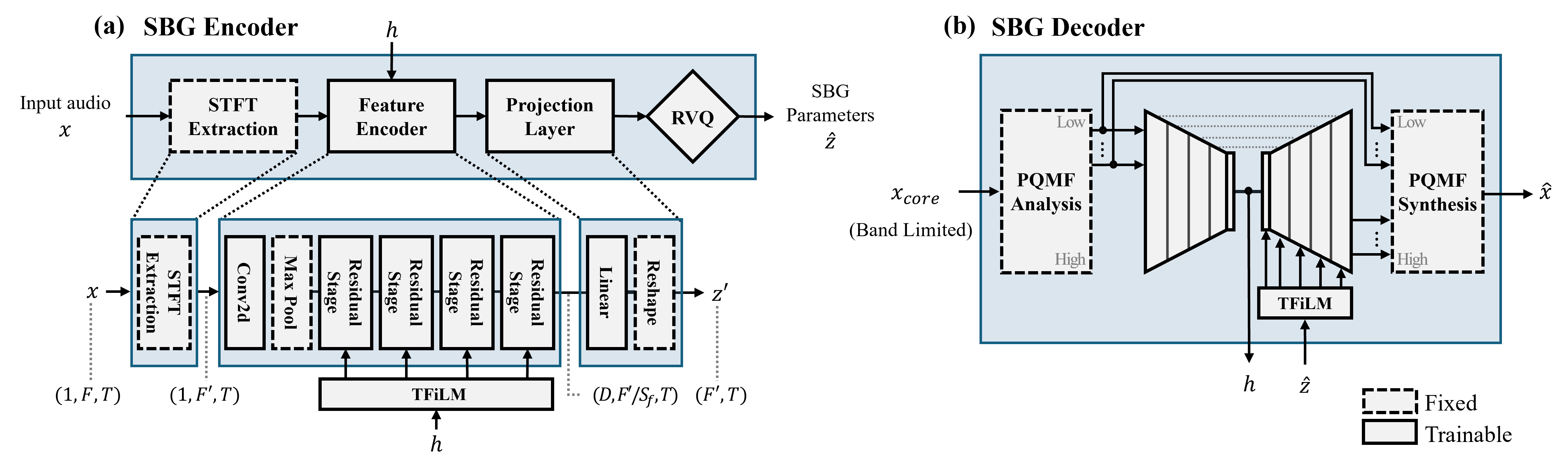}
    \label{fig:fig}
    \vspace{-0.5cm}
    \caption{Detailed descriptions of the encoder and the decoder architecture: (a) SBG encoder architecture with details of the feature encoder and the projection layer, (b) SBG decoder architecture.}
    \label{fig2}
    \vspace{-0.5cm}
\end{figure*}
% learnable, fixed 구분 (가능하면 색으로), label 달아주기
% SBG Encoder: Residual Unit -> Residual Stage
% FiLM -> TFiLM

% \begin{figure*}[t]
%     \centering
%     \includegraphics
%     [width=\linewidth]
%     % {figures/fig3-t3.png}
%     % y축 crop 한 그림
%     {figures/fig3.png}
%     \vspace{-0.7cm}
%     \caption{Rate-distortion curves comparing FDK HE-AAC v1 and the proposed model, both generated using the AAC core output at the same side information bitrate. Distortion measure: (a) NMR $\downarrow$ (b) 2f-model MMS $\uparrow$ (c) ViSQOL MOS $\uparrow$}
%     \label{fig:fig3}
%     \vspace{-0.1cm}
% \end{figure*}

\section{Proposed Method}
The proposed method consists of two main components---SBG encoder and decoder---as shown in Figure \ref{fig2}.
The SBG encoder extracts SBG parameters (lying in an embedding space) from the STFT coefficients of the input audio, while the SBG decoder generates high-frequency subband signals for Pseudo-Quadrature Mirror Filters (PQMF)~\cite{nussbaumer1984pqmf} based on the transmitted SBG parameters and the decoded low-frequency subband signals from the core-codec.
% ======= Added ==========
% We also introduce an auxiliary encoder, which shares its architecture with the SBG decoder’s bottleneck. This auxiliary module provides additional conditioning based on the core-band signal.
Additionally, the output embeddings from the bottleneck of the SBG decoder are provided to the SBG encoder as a conditional input, making the extraction of side information dependent on the coded core bands.
% to enable a core band-dependent side information extraction
% Detailed description for the conditioning process is provided in Section \ref{sub_cond}.
% ======= Added ==========
Compared to SBR, our method replaces the rule-based parameter extraction and synthesis pipeline with a fully neural-network-based approach, optimized in an end-to-end manner.
% both the parameter extraction and high-frequency generation processes in n-SBG are implemented using neural networks and jointly optimized in an end-to-end manner.
The details of each processing module are explained in the following subsections.

\subsection{SBG encoder}
\vspace{-2.5pt}
%% ---> Residual Stage name ?
%% ----> Basic building block ?
The SBG encoder is designed with an architecture similar to ResNet-18~\cite{he2016res}.
% Transformaition (STFT)
The input audio signal $x \in \mathbb{R}^{T'}$ is transformed into a log-power spectrogram $x_{\text{stft}} \in \mathbb{R}^{1 \times F \times T}$, where $T'$, $T$, and $F$ indicates the length of the input audio signal, the number of frames, and the number of frequency bins, respectively. 
Only the frequency bins corresponding to the range generated by the SBG decoder are provided to the feature encoder.
The selected spectral coefficients, represented as $\hat{x} _{\text{stft}}\in \mathbb{R}^{1 \times F' \times T}$, are then processed by the feature encoder.

% ======= Added ==========
%% ==> SBG decoder's Encoder?
% \vspace{4pt}
% \noindent\textbf{Auxiliary encoder.} To further refine feature extraction, the SBG encoder incorporates an auxiliary encoder, sharing both its architecture and weights with the SBG decoder’s encoder. This module extracts a latent representation \(h\) from the core-band output, which is fused with the spectral features $\hat{x} _{\text{stft}}$. 
% The detailed description of the conditioning process is provided in Section 3.2.
% By integrating both sources of information, the encoder ensures that the quantized parameter $\hat{z}$ is optimally conditioned for high-frequency reconstruction.
% ======= Added ==========

% -------- Details of Feature Encoder
% \vspace{4pt}
% \noindent\textbf{Feature extraction.} 
The feature encoder (shown in Figure \ref{fig2}.a) consists of an initial 2D-convolution with a kernel size of $(7,7)$, stride factor of $(2,1)$ and an output channel size of $D/8$, followed by max pooling layer with a kernel size of $(3,3)$ and stride factor of $(2,1)$, and four \textit{Residual Stages}.
Each \textit{Residual Stage} contains $3 \times 3$ convolutions and ReLU activations.
The last three stages progressively double the input channel size and halve the frequency resolution.
%Additionally, all residual stages are conditioned with a latent embedding $h$ from the SBG decoder, so that the SBG encoder have prior knowledge about the core codec output.
The feature encoder then produces an output embedding $z \in \mathbb{R}^{D \times (F'/ S_f) \times T}$, where $S_f = 32$.
All convolutional layers in the feature encoder are causal.
% Projection Layer
The extracted feature embedding $z$ is linearly projected into an $S_f$-dimensional space, and then reshaped into $z'\in \mathbb{R}^{F' \times T}$.

%---------- RVQ Layer
% \vspace{4pt}
% \noindent\textbf{Residual Vector Quantization.} 
Following the quantization scheme of DAC~\cite{kumar2024DAC}, we use RVQ to quantize $z'$ into SBG parameter $\hat{z}$.
The RVQ consists of $N_q$ layers of vector quantization (VQ), each with an $N$-dimensional codebook containing $M$ learnable code vectors, where $N < F'$.
% Single stage
% At each stage $q \in \{1, \ldots, N_q\}$, the $F'$-dimensional input vector is projected into an $N$-dimensional space and mapped to the code vector with the highest cosine similarity.
% The selected code vector is then projected back into a $F'$-dimensional output vector. The residual between the input and output vectors is passed to the following VQ layer.
% After all $N_q$ stages, the sum of the outputs from all VQ layers forms an additional parameter $\hat{z}$.
% Bitrate
The bit-rate of the SBG parameter is calculated as $\frac{f_s}{H} \cdot N_q \cdot \lceil \log_2 M \rceil (\textrm{bps})$,
%\frac{1}{1000} \cdot \frac{f_s}{H} \cdot N_q \cdot \log_2(M),
where $f_s$ denotes the sampling rate and $H$ is the hop length of the STFT.

% ================================
\subsection{SBG decoder}
\vspace{-2.5pt}
As shown in Figure \ref{fig2}.b, the SBG decoder receives two inputs: (1) the coded core-band signal $x_{\text{core}}$, which is decomposed into critically-sampled subbands $(b_{1}^{(\text{core})}, \ldots, b_{N_{\text{core}}}^{(\text{core})})$ and (2) the SBG parameters $\hat{z}$ from the SBG encoder.
% Using these, the SBG decoder generates high-frequency bands $(b_{N_{\text{core}}+1}^{(gen)}, \ldots, b_{N_{\text{core}} + N_{\text{HF}}}^{(gen)})$, which are then synthesized with the coded core-bands into the bandwidth-extended output $\hat{x}$.
%==========
Using these, the SBG decoder generates high-frequency bands $(b_{N_{\text{core}}+1}^{(gen)}, \ldots, b_{N_{\text{core}} + N_{\text{HF}}}^{(gen)})$, which are subsequently combined with the subbands of the coded core-bands and synthesized into the bandwidth-extended output $\hat{x}$.
%==========
We use 32-channel PQMF filterbanks for the subband analysis and synthesis.

%\vspace{4pt}
%\noindent\textbf{BWE module.}
We adopt SEANet~\cite{tagliasacchi2020seanet, li2021StreamingSEANEt}, a time-domain speech BWE model, as the backbone for the SBG decoder and extend its architecture to process multi-channel input and output.
% ==========  modifications  =========================
% ========== writing in an sequential order ==========
Specifically, the SBG decoder begins with an initial 1D-convolution layer that expand the channel size from $N_{\text{core}}$ to $C$.
% Encoder Block
Next, the feature maps pass through four \textit{Encoder Blocks} (i.e., Embedding Extractor), which downsample along the temporal axis and double the channel size. 
%Each \textit{Encoder Block} comprises three \textit{Residual Units} and a strided convolution, so that the temporal resolution is progressively downsampled while the number of channels is doubled.
% Bottle neck
Following the \textit{Encoder Blocks}, the signal reaches the bottleneck stage, which consists of two 1D-convolution layers.
The first layer reduces the channel size from $16C$ into $4C$, and the second layer expands it back to $16C$.
Then, four \textit{Decoder Blocks} (i.e., Band Generator) successively invert the process of the \textit{Encoder Block}.
An addition-based skip connection links each corresponding encoder and decoder block.
Finally, the last 1D-convolution outputs $N_{\text{HF}}$ channels corresponding to the high-frequency subbands to be reconstructed. 
We set the stride factors of the downsampling layers to $(1,2,2,2)$.
Given that the PQMF analysis process already downsamples the input signal by a factor of $32$, the downsampling layers further reduces the time resolution by an additional factor of $2^3=8$, resulting in an overall reduction up to $256$.
All convolution layers within the SBG decoder are causal, ensuring real-time processing that remains consistent with the causal structure of the SBG encoder.

\subsection{Conditioning scheme} \label{sub_cond}
\vspace{-2.5pt}
The n-SBG leverages two complementary forms of conditioning.
First, the SBG decoder extracts latent embeddings $h \in \mathbb{R}^{4C \times (T' / 256)}$ from its bottleneck and feeds it to the four \textit{Residual Stages} of the SBG encoder.
This makes the extraction of SBG parameters dependent on the coded core band, resulting in more efficient usage of the bit-rate.
% This additional information helps the SBG encoder extract SBG parameters $\hat{z} \in \mathbb{R}^{S_f \times (T' / H)}$ that are more aligned with the high-frequency generation process in the decoder.
% Specifically, this conditioning is applied in the SBG encoder via Feature-wise Linear Modulation (FiLM) \cite{perez2018film}, where \(\hat{h}\) modulates the activations of four \textit{Residual Units}.  
Second, the SBG parameter $\hat{z} \in \mathbb{R}^{S_f \times (T' / H)}$ is provided to the last bottleneck layer and the four \textit{Decoder Blocks} of the SBG decoder, facilitating more accurate reconstruction of high-frequency bands.
% Second, the SBG decoder utilizes \(\hat{z}\) to modulate the creation of high-frequency subbands.
%This conditioning is applied after the BWE module’s bottleneck, influencing the subsequent convolutional layers responsible for high-frequency synthesis.

% Method for conditioning
% We adopt Temporal Feature-wise Linear Modulation~\cite{birnbaum2019temporal} for a conditioning method to modulate each timestep independently.
We utilize Temporal Feature-wise Linear Modulation (TFiLM)~\cite{birnbaum2019temporal} to modulate the activation of a specific layer at each timestep, conditioning it based on a conditional input. 
Let $a \in \mathbb{R}^{C_a \times \ldots \times T_a}$ be the activation to be modulated, where $C_a$ is the channel size and $T_a$ is the number of timesteps.
%, omitting the batch dimension.
We extract two parameters $\beta, \gamma \in \mathbb{R}^{C_a \times T_a}$ from a conditional input $b \in \mathbb{R}^{C_b \times T_b}$.
If $T_b > T_a$, we use a strided convolution; otherwise, we replicate each timestep of $b$ to align the temporal resolution with $a$.
% FiLM-generator
The resampled tensor $b_{\text{re}}\in \mathbb{R}^{C_b \times T_a}$ is then mapped to the shape $(C_a, T_a)$ through a point-wise linear projection.
Finally, each timestep of the activation $a$ is modulated as follows:
\begin{equation}
a^{(\text{mod})}_t = \gamma'_t \cdot a_t + \beta'_t, \quad \text{for  } 0 \leq t < T_a,
\end{equation}
where $a_t, a^{(\text{mod})}_t \in \mathbb{R}^{C_a \times \cdots \times 1}$, and $\beta_t, \gamma_t \in \mathbb{R}^{C_a \times 1}$ are reshaped into $\beta'_t \in \mathbb{R}^{C_a \times \cdots \times 1}$ and $\gamma'_t \in \mathbb{R}^{C_a}$.
% where $a_t, a^{(\text{mod})}_t, \beta_t \in \mathbb{R}^{C_a \times \cdots \times 1}$ and $\gamma_t \in \mathbb{R}^{C_a}$.

% \begin{figure*}[t]
%     \centering
%     \begin{subfigure}[b]{0.33\textwidth}  
%         \centering
%         \includegraphics
%         [width=\linewidth]
%         {figures/fig3-1.png}
%         \caption{}
%         \label{fig:fig3a}
%     \end{subfigure}
%     \hfill 
%     \begin{subfigure}[b]{0.33\textwidth} 
%         \centering
%         \includegraphics
%         [width=\linewidth]
%         {figures/fig3-2.png}
%         \caption{}
%         \label{fig:fig3b}
%     \end{subfigure}
%     \hfill
%     \begin{subfigure}[b]{0.33\textwidth} 
%         \centering
%         \includegraphics
%         [width=\linewidth]
%         {figures/fig3-3.png}
%         \caption{}
%         \label{fig:fig3c}
%     \end{subfigure}
%     \caption{Rate-distortion curves comparing FDK HE-AAC v1 and the proposed model, both generated using the AAC core output at the same side information bitrate. Distortion measure: (a) NMR $\downarrow$ (b) 2f-model MMS $\uparrow$ (c) ViSQOL MOS $\uparrow$}
%     \label{fig2}
% \end{figure*}

% Kang: Core AAC-LC와 Blind BWE를 포함시켜야 하나? 차이가 워낙 많이 나서 제안 방법과 SBR차이를 정확히 보기 어려움. 굳이 넣고 싶다면 Ablabtion study 형태로 본문 문장에서 한 두줄 포함시키는 것이 좋을 것 같음.
\begin{figure*}[t]
    \centering
    \includegraphics
    [trim={0 10pt 0 5pt}, clip, width=\linewidth]
    {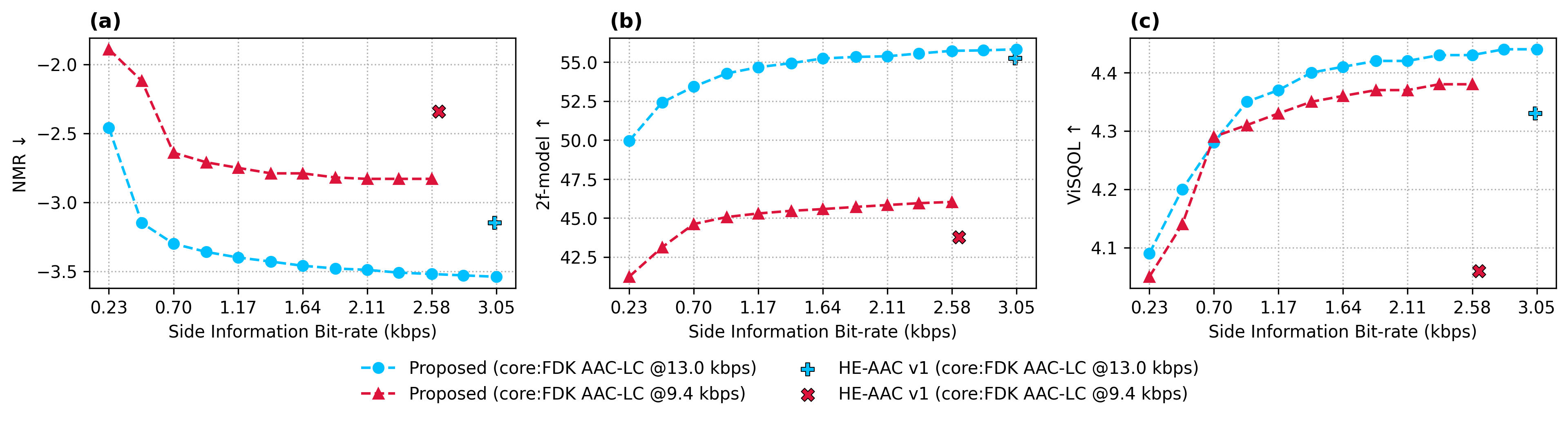}
    % y축 crop 한 그림
    % {figures/fig3.png}
    \vspace{-0.5cm}
    \caption{Rate-distortion curves comparing FDK HE-AAC v1 and the proposed model, both generated using the AAC core output at the same side information bitrate. Distortion measure: (a) NMR $\downarrow$ (b) 2f-model MMS $\uparrow$ (c) ViSQOL MOS $\uparrow$}
    \label{fig:fig3}
    \vspace{-0.4cm}
\end{figure*}

\section{Experiments}
\subsection{Training Objectives}
\vspace{-2.5pt}
%Our training objective is primarily inspired by \cite{kumar2024DAC}. While prior multi-band synthesis approaches \cite{yang2021mbmelgan} mainly utilize subband reconstruction loss, we found that applying full-band reconstruction loss was sufficient for effective training.
We apply a GAN framework~\cite{zeghidour2021soundstream, defossez2022encodec, kumar2024DAC} to train the n-SBG, treating the entire framework as a generator.
The training objective of the generator ($\mathcal{L}^{total}_G$) is composed of a weighted sum of various loss functions:
\begin{equation}
    \begin{aligned}
        \mathcal{L}^{total}_G &= \lambda_{mel} \mathcal{L}_{mel} 
        + \lambda_{adv} \mathcal{L}^{adv}_G 
        + \lambda_{fm} \mathcal{L}_{fm} \\
        &\quad + \lambda_{cb} \mathcal{L}_{cb} 
        + \lambda_{cm} \mathcal{L}_{cm},
    \end{aligned}
\end{equation}
where the weighting coefficients are set as 
\(\lambda_{mel} = 15\), 
\(\lambda^{adv}_G  = 3\), 
\(\lambda_{fm} = 6\), 
\(\lambda_{cb} = 1\), and 
\(\lambda_{cm} = 0.5\).
The multi-scale mel-reconstruction loss, $\mathcal{L}_{mel}$, is computed across seven different frequency resolutions~\cite{kumar2024DAC} and is defined as follows:
% Following the prior work \cite{kumar2024DAC}, we adopt a multi-scale mel-reconstruction loss ($\mathcal{L}_{mel}$) calculated across different resolutions:
\begin{equation}
    \mathcal{L}_{mel} = \sum_{i=1}^{7} \parallel \log_{10} M_{i}(x) - \log_{10} M_{i}(\hat{x}) \parallel_1,
\end{equation}
where \( M_i(\cdot) \) represents the mel-spectrogram at scale $i$, computed using a window length of $2^{4+i}$, a hop size of $2^{2+i}$, and $5 \times 2^i$ mel bins.
For adversarial training, we use the hinge loss ($\mathcal{L}^{adv}_G $)~\cite{lim2017geometric} and the feature matching loss ($\mathcal{L}_{fm}$).
% \begin{equation}
%     \mathcal{L}_{G}^{adv} = - \mathbb{E}_{\hat{x}} [D(\hat{x})],
% \end{equation}
% \begin{equation}
%     \mathcal{L}_{fm} = \sum_{l} || D_l(x) - D_l(\hat{x}) ||_1,
% \end{equation}
% where \( D_l(\cdot) \) represents feature maps from the \( l \)-th discriminator layer.
Additionally, we use two auxiliary losses to train the RVQ: the codebook loss ($\mathcal{L}_{cb}$) and the commitment loss ($\mathcal{L}_{cm}$)~\cite{van2017vqvae}.
% :
% \begin{equation}
%     \mathcal{L}_{cb} = || sg[z_e] - z_q ||^2, \quad
%     \mathcal{L}_{cm} = || z_e - sg[z_q] ||^2,
% \end{equation}
% where \( sg[\cdot] \) denotes the stop-gradient operation, \( z_e \) represents the encoder output, and \( z_q \) is the quantized vector obtained from the codebook.\
We employ multi-band STFT-based discriminators~\cite{kumar2024DAC} and multi-period discriminators~\cite{kong2020hifigan} to enable high-fidelity reconstruction. These discriminators are trained using a hinge loss ($\mathcal{L}_{D}^{adv}$) with a weighting factor of $\lambda^{adv}_D = 1$.
% \begin{equation}
%     \mathcal{L}_{D}^{adv} = \mathbb{E}_{x} [\max(0, 1 - D(x))] 
%     + \mathbb{E}_{\hat{x}} [\max(0, 1 + D(\hat{x}))].
% \end{equation}

% Target signal
All training objective functions are calculated between the output signal $\hat{x}$ and the target signal $x_{\text{tgt}}$, where $x_{\text{tgt}}$ is synthesized from $(b_{1}^{(core)}, \ldots, b_{N_{\text{core}}}^{(core)}, b_{N_{\text{core}}+1}^{(input)}, \ldots, b_{N_{\text{core}} + N_{\text{HF}}}^{(input)})$ via a PQMF synthesis filterbank.
%, where $b_{i}^{(input)}$ denotes $i$-th PQMF subbands from the input signal $x$.
% The target signal serves as a reference for calculating training objectives, representing the ideal high-frequency content.
% It is obtained by applying PQMF analysis to the original uncompressed audio before encoding, extracting the \(N_{\text{HF}}\) high-frequency bands.  
% These bands are then combined with the \(N_{\text{core}}\) bands extracted from the core codec output, forming the final reference signal for high-frequency reconstruction.

\subsection{Dataset and Experimental Setup}
\vspace{-2.5pt}
% Dataset
% ===> Test dataset?
% We use three datasets for training: FSD-50K~\cite{fonseca2021fsd50k}, a general audio dataset containing 51,000 samples; MUSDB-18~\cite{rafii2017musdb18}, a music dataset with various instrument and vocal tracks; and VCTK~\cite{veaux2017vctk}, a speech dataset featuring recordings from over 100 speakers. All datasets have a sampling rate of 48 kHz.
% We use three datasets for training: FSD-50K~\cite{fonseca2021fsd50k} for general audio, MUSDB18-HQ~\cite{rafii2017musdb18} for music, and 
% VCTK~\cite{veaux2017vctk} for speech. 
For training, we utilize three datasets: FSD-50K~\cite{fonseca2021fsd50k}, MUSDB18-HQ~\cite{rafii2017musdb18}, and VCTK~\cite{veaux2017vctk}. These datasets provide approximately 70 hours of diverse sound events, 30 hours of music, and 30 hours of speech content, respectively.
For evaluation, we used 43 candidate test items for USAC standardization~\cite{neuendorf2009usac}, consisting of 11 for speech, 18 for music, and 14 for mixed signals, respectively. All datasets have a sampling rate of 48 kHz.

% Baseline codec
In the experiment, we compare n-SBG with Fraunhofer FDK HE-AAC v1\footnote{\url{https://tsrac.ffmpeg.org/wiki/Encode/AAC}}, where both share the same core codec.
We used HE-AAC v1 at 12 and 16 kbps bitrate settings. The split of the subbands $(N_{\text{core}}, N_{\text{HF}})$ for the n-SBG follows the configuration of HE-AAC v1, using $(5, 10)$ and $(5, 11)$ at 12 and 16 kbps respectively.
The n-SBG encoder outputs a 512-dimensional vector per STFT frame, with both window length and hop length set to 2048, i.e., \( (D,H)=(512,2048)\).
RVQ of n-SBG encoder has a maximum bit-rate similar to that of the SBR in HE-AAC v1.
Each codebook contains 1024 number of 8-dimensional code vectors, where $N_q=11$, $13$ for 12 and 16 kbps setups, respectively.
% Codebook small
For inference, the bit-rate of the SBG parameters can be adjusted by bypassing the last few VQ layers.
The n-SBG decoder has a channel size of $C=64$.
% % Optimizer
We trained n-SBG using the Adam optimizer with an initial learning rate of \(1.0 \times 10^{-4}\), \((\beta_1, \beta_2) = (0.5, 0.9)\), and exponential learning rate decay with $\gamma = 0.999996$ for both the generator and the discriminators.

\subsection{Experimental Results}
\vspace{-2.5pt}
% \begin{figure}[t!]
% % \flushleft
% % \hfill
% % \vspace{0.1cm}
% \centering
% \includegraphics[width=\columnwidth]{figures/fig4.png}
%   \vspace{-0.6cm}
%   \caption{ABX test results}
%   \label{fig:fig4}
%   \vspace{-0.3cm}
% \end{figure}

% ======
% 가독성이 떨어지는데 ylabel을 어떻게 할지 고민해봐야 함
% ======

\begin{figure}[t]
    \centering
    \vspace{-8pt} 
    \begin{subfigure}[b]{\linewidth}
        \centering
        \includegraphics[trim={0 24pt 0 3pt}, clip, width=\linewidth]{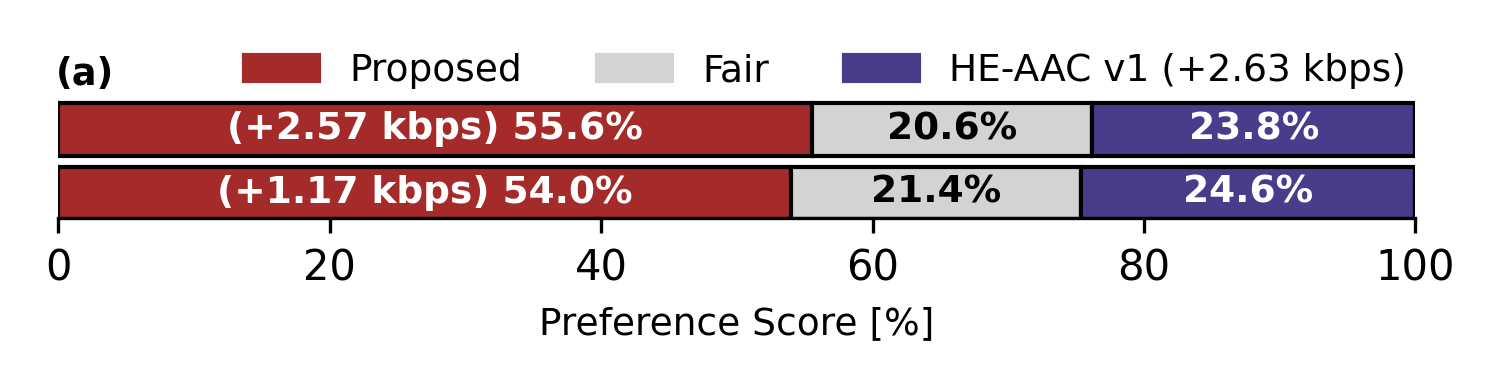}
        \vspace{-5pt}
        % \caption{Using AAC-LC @9.6 kbps with varying side information bit-rates.
        % % The proposed model is evaluated with different side-information bit-rate usage 
        % \newline
        % (Top: similar to HE-AAC v1, Bottom: approximately half).}
        \label{fig:fig4a}
    \end{subfigure}
    %\vspace{-2pt} 
    \begin{subfigure}[b]{\linewidth}
        \centering
        \includegraphics[trim={0 12pt 0 3pt}, clip, width=\linewidth]{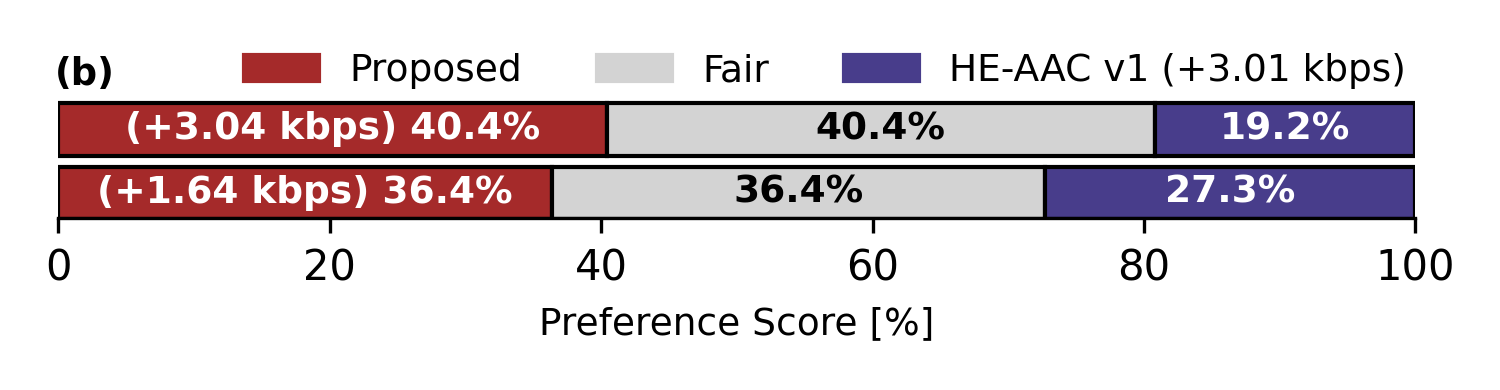}
        \vspace{-10pt}
        % \caption{Using AAC-LC @13.0 kbps with varying side information bit-rates.}
        \label{fig:fig4b}
    \end{subfigure}
    
    \vspace{-8pt} 
    \caption{A/B preference test results comparing n-SBG with HE-AAC v1 under different core bit-rate conditions: (a) AAC-LC @9.4 kbps (b) AAC-LC @13.0 kbps. Values in the parentheses indicate the bit-rate of side information.}
    \label{fig:fig4}
    \vspace{-15pt}
\end{figure}

% \begin{figure}[t]
%     \centering
%     \vspace{-8pt} 
%         \centering
%         \includegraphics[width=\linewidth]{figures/fig4-c.png}
%         \vspace{-15pt}
%         % \caption{Using AAC-LC @9.6 kbps with varying side information bit-rates.
%     \caption{Average preference score (\%) for ABX tests comparing the proposed method with HE-AAC v1 under different core bit-rate conditions. Proposed method utilizes approximately half the bit-rate for side information: (Top) AAC-LC @9.6 kbps, (Bottom) AAC-LC @13.0 kbps.}
%     \label{fig:fig4}
%     \vspace{-10pt}
% \end{figure}

% (Ablation) w\o core conditioning 과 proposed n-SBG 비교.
% Metric 별 경향성 설명 추가하기 (중요한 내용만)
% Data type 별로 mertric 보여주고 분석하기
% 자리 남으면 Spectrogram 추가하기 (넣을거면 objective & subjective eval 하나의 subsection으로 합치기)
We evaluate the performance of the proposed method using NMR~\cite{brandenburg1992nmr}, 2f-model MMS~\cite{kastner20192f}, and ViSQOL MOS (audio mode)~\cite{chinen2020visqol} as objective metrics. 
Figure \ref{fig:fig3} shows the rate-distortion curves based on the bit-rate of the side information.
We compare n-SBG and HE-AAC v1 at various bit-rates for the side information, both utilizing AAC-LC at either 13.0 kbps or 9.4 kbps as a core codec.
% Reference signal
For reference signals, full-band signals were low-pass filtered to match the bandwidth specifications of HE-AAC v1.
% Similar bit comparison
When consuming similar bit-rates for the side information, the n-SBG consistently outperforms the HE-AAC v1 across all objective metrics.
% Half bit comparison
% Furthermore, by adjusting the bit-rate of the RVQ at the inference stage, the n-SBG achieves comparable performance while using approximately half bit-rates for additional parameters compared to the SBR in HE-AAC v1, in terms of 2f-model MMS.
With respect to 2f-model MMS, n-SBG maintains comparable performance to the HE-AAC v1 while utilizing approximately half the bit-rate for the side information.

% \subsection{Subjective Evaluations}
% =====> P-value 계산 가능하면 해볼 수도
Figure~\ref{fig:fig4} illustrates the results of an A/B preference test conducted with 14 participants,
using randomly selected 9 audio samples (3 speech, 3 music, and 3 mixed) from the test set.
The listening test follows the same experimental setup as the objective evaluation.
%, except that the n-SBC consumes approximately half the bit-rate for side information compared to the SBR in HE-AAC v1.
%but compares the model's performance when using a lower side information bit-rate than HE-AAC v1. % half bit setup
% where the AAC core codec operates at either 13.0 kbps or 9.4 kbps.
% =====> Same bit
% When the similar bit-rate is allocated for the side information, the n-SBG is consistently preferred over the HE-AAC v1.
% =====> half bit
% Notably, even when the bit-rate allocated for side information is nearly halved, the proposed model still achieves higher preference scores compared to HE-AAC v1.
% Results show that even with less side information, the proposed model achieves higher preference scores compared to the HE-AAC v1.
Results show that n-SBG achieves higher preference scores compared to HE-AAC v1, even when the n-SBG allocates about half the bit-rates for the side information.
Furthermore, the preference for n-SBG over SBR becomes more dominant as the bit-rate of the core codec decreases, indicating the effectiveness of n-SBG in low bit-rate conditions.
% Furthermore, the higher preference for signals with fewer bits allocated to the core indicates that the proposed model achieves better bit efficiency under harsh conditions.
% These results demonstrate that replacing a traditional rule-based module with a DNN-based modules enables a significant reduction in bit-rate for additional parameters while maintaining audio quality, especially in low bit-rate conditions.

% ======> 오디오 신호 별 경향성 및 limitation 언급
\def\arraystretch{0.8} % 글 쓸 자리 부족하면 table 높이 줄이기
\begin{table}[t]
    \centering
    \caption{Objective scores comparing Blind SBG and n-SBG}
    \vspace{-8pt}
    %\footnotesize
    \resizebox{\linewidth}{!}{%
    \begin{tabular}{lccc}
    \toprule
    \textbf{}   & \textbf{NMR $\downarrow$} & \textbf{2f-model $\uparrow$} & \textbf{ViSQOL $\uparrow$} \\ 
    \midrule
    Core: AAC-LC @ 9.4 kbps & -1.74 & 28.34 & 2.52 \\
    \hspace{2.1em} + Blind SBG          &  3.78  & 32.15 & 3.38 \\ 
    \hspace{2.1em} + n-SBG (+2.57 kbps)  & -2.80  & 46.04 & 4.38 \\ 
    \midrule
    Core: AAC-LC @13.0 kbps  & -2.05 & 33.00 & 2.38 \\
    \hspace{2.1em} + Blind SBG          &  1.51  & 37.47 & 3.34 \\ 
    \hspace{2.1em} + n-SBG (+3.04 kbps)  & -3.54  & 55.83 & 4.44 \\ 
    \bottomrule
    \end{tabular}
    }
    \label{table1}
    \vspace{-0.6cm}
\end{table}
\def\arraystretch{1.0}

% \subsection{Ablation and Discussion}
\subsection{Ablation Study and Discussion}
\vspace{-2.5pt}
% =====> Blind, Core에 대한 실험 결과 언급
% bandwidth-limited core signals?
In Table \ref{table1}, we compare the objective evaluation results of the core codec output, the n-SBG output, and the output from n-SBG without side information, referred to as Blind SBG.
The Blind SBG significantly underperforms n-SBG and shows even worse NMR than the core codec output, indicating severe audible noise is introduced.
These results highlight the necessity of side information for generating high-frequency components of complex audio signals.
% In Table \ref{table1}, we present the objective metric scores for bandwidth-limited core signals and blind bandwidth-extended (BWE) signals, which generate high-frequency components solely from the core codec output without utilizing SBG parameters. 
% The results indicate that replacing SBR with a blind BWE model is not suitable for complex audio signals, as it struggles to reconstruct high-frequency components accurately.

% Discussion
Despite the overall performance of n-SBG surpassing SBR, its effectiveness varies depending on the characteristics of audio signals.
n-SBG generates more realistic and vibrant transient components than the SBR but often struggles with generating tonal components with prominent harmonic structures.
For future research, we will explore more effective conditioning strategies and high-frequency generation methods while also aiming to develop a system adaptive to various core codecs with different bit-rates.

% While our model aims to reconstruct high-frequency content in general audio at 48 kHz, its performance varies across audio types.
% In transient-rich musical signals, the model captures diverse spectral characteristics more effectively than SBR.
% However, for tonal sounds with prominent harmonic structures, its generalization capability remains limited, likely due to insufficient spectral resolution in the high-frequency range.
% Future research will refine the conditioning strategy and model architecture to improve frequency resolution and explore broader integration with existing codecs.
% focus on refining the model’s conditioning strategy and architectural design to enhance frequency resolution and adaptability, while also exploring its integration with various existing codecs.
% Future research will focus on developing a unified architecture capable of adapting to various bitrate constraints.
% \vspace{-2.5pt}
\section{Conclusion}
In this work, we propose neural Spectral Band Generation (n-SBG), an alternative approach to rule-based Spectral Band Replication (SBR).  
The proposed method reconstructs high-frequency components by encoding and transmitting parametric information, which the SBG decoder efficiently utilizes for generation.  
Experimental results demonstrate that the n-SBG significantly outperforms the SBR at comparable bit-rates, with particularly notable efficiency gains observed in low bit-rate scenarios.
% Furthermore, our framework can be easily integrated into various audio coding pipelines, since both the SBG encoder and decoder operate on the multi-band signals.
However, n-SBG struggles with generating complex tonal components and requires training separate models for different bit-rates. Developing a unified system adaptive to various core codecs and bit-rates should be investigated in future works.

%%%%%%%%%%%%%%%%%%%%%%%%%%%%%%
\newpage
\bibliographystyle{IEEEtran}
\bibliography{refs}

\end{document}